# AERODYNAMIC MODELS FOR HURRICANES

# I. Model description and horizontal motion of hurricane

A.I. Leonov

The University of Akron, Akron, OH 44325-0301

E-mail: **leonov@uakron.edu**

**Abstract**

Aerodynamic models are developed to describe coherent structures and transport processes in hurricanes moving over open seas. The models consist of the lower boundary layer and upper adiabatic layer. Except friction at the air/sea interface, proposed modeling avoids the common turbulent approximations while using explicitly or implicitly basic stability constraints. The models analyze dynamics of upper hurricane adiabatic layer, dynamics and transport processes in hurricane boundary layer, and genesis and maturing of hurricane. The proposed modeling provides a rude enough but consistent analytical description of basic processes in hurricanes.

The present paper qualitatively describes the model of mature hurricane, briefly discusses the basic thermodynamic relations and aerodynamic equations, and establishes the principles of horizontal motion for mature hurricane.

## 1.1. Introduction: Background and scope of paper series

Intense tropical cyclones (TC's), called hurricanes in Atlantics or typhoons in Pacific were extensively investigated in the recent 50-60 years. Many details of the TC's have been observed and experimentally studied by concert efforts of various teams with the use of aircrafts, buoys, ships, and satellites. These observations and related models created a detailed qualitative picture of hurricane structure, documented in several texts [1-6]). They also show that hurricane could be viewed as a complicated natural thermo-aerodynamic machine, which works with surprising efficiency and stability.

Several works were also aimed at elaborating conceptual, idealized models, which could properly describe some hurricane features and their effect on hurricane severity. E.g. papers [7-9], based on a Carnot model with friction evaluated the potential intensity of hurricanes. Paper [10] used another approach to the same problem. Paper [11] also evaluated hurricane strength. Possible effects of dissipative heating on hurricane intensity in the boundary layer near the maximal wind were discussed in paper [12]. More complicated physical processes related to the dynamic role of hurricane mesovortices in the hurricane eye were recently experimentally modeled in laboratory and discussed in paper [13]. A theory of ocean spray thermodynamics was developed by Lighthill [14], and the spray effect on the dynamics of near water air turbulence in hurricane eye wall (EW) was elaborated in paper [15]. Highly coupled interactions between the turbulent wind and oceanic waves near the air/sea interface have been thoroughly investigated and compared with observations in many research papers, well cited and well presented in the Chapters 3.5 and 3.9 of the monograph [16]. These results have not, however, applied to hurricanes. Other conceptual ideas were scattered throughout various papers being organically mixed with numerical schemes and their implementations. Here one could mention models by Ooyama [17,18] and Emanuel [19,20] investigated the intriguing aspect of hurricane maturing.

Many other papers attempted to describe the observed hurricane phenomena, using turbulent baroclinic and barotropic numerical models. For more detailed description, these models involve both the thermodynamic and some empirical relations describing the cloud dynamics, evaporation, hailing etc. Along with physical parameters, these models contain plenty of adjustable parameters describing the turbulent diffusivity coefficients. Nevertheless, after adjusting these parameters to recorded observations of a past hurricane, some numerical multi-scale models were capable to incorporate many physical phenomena and using fine finite elements describe the recorded hurricane structure and its evolution reasonably well (e.g. see paper [21] and references there). Some models also capable to analyze the effect of variability of $\beta$ - projection of Earth angular velocity within the hurricane path and even within the region of hurricane itself. Most importantly for hurricane forecasts, these models attempted to predict the horizontal hurricane motion by postulating that hurricanes move to the direction of maximal potential vorticity, which captures both the temperature and pressure effects. In order to make a forecast of hurricane motion one should know the environmental situation at least in the synoptic scales. Therefore these models are often supposed to interact with the current synoptic and lower scale observations. Two recent extensive reviews of the conceptual and numerical models could be found in Refs. [22, 23].

In spite of many impressive results, there are several unresolved fundamental problems in understanding of hurricane functioning, propagation and genesis:

- Current, mostly numerical, models of hurricanes neither properly exposed nor formalized the central roles of hurricane eye wall in the dynamics and transport processes in hurricanes, not mentioning a clear description of their vertical/radial structure. The surface pressure distribution, one of the important measured factors in hurricanes, was not properly described too.
- It is still poorly known what the main physical processes in the hurricane boundary layer (HBL) are responsible for maintaining the stability of hurricane, as well as the reason for surprisingly growing angular momentum in the radial direction in HBL.
- Still unknown are the reasons for hurricane rotation in cyclonic direction near the hurricane bottom and anti-cyclonic in the upper layer.
- Still unknown is also a reason (and description) of why radial flow in hurricane changes directions with altitude, from inward in lower layer to outward in upper layer.
- The physics involved in the hurricane genesis and maturing is still vague.

The main objective of current series of papers is to resolve the above problems by developing and analyzing quantitative conceptual and prognostic idealistic models. These models being rude enough provide a consistent, albeit analytical description of the basic structure and physical phenomena in hurricanes. Based on observations, the main conceptual idea in the series is that the hurricanes are *coherent structures*. Therefore the proposed and developed models mostly avoid the common turbulent approximations which include unknown turbulent diffusivities, except friction factors at the air/sea interface. The developed large scale approach does not deny the importance of turbulence in the hurricane description but rather attempts to localize the turbulent effects to special, lower scale regions. This is quite similar to many other branches of geophysical fluid mechanics. The modeling, based on aerodynamics of ideal gas, requires as much as possible implementing continuity for various dynamic variables. This "principle of maximal continuity" is necessary to avoid the Kelvin-Helmholtz (K-H) instability. In this regard, we consider the small areas where the aerodynamic models cannot be continuous as the subject of the K-H instability with a quick turbulization in these zones. Therefore the stability constraints are explicitly or implicitly used in our modeling.

This first paper in the series qualitatively schematizes a simplified picture of hurricanes, and discusses the principles of horizontal motion of hurricanes. The second paper analyzes the dynamic processes in the upper hurricane layer, and describes the structure of the layer. The third paper analyzes basic dynamic and thermodynamic processes and balance relations in the hurricane boundary layer, located between the ocean interface and the upper layer. The fourth paper presents a simple analytical model for hurricane genesis and maturing, and briefly discusses the similar problems for tornados.

## 1.2. Qualitative description of model structure and basic processes in mature hurricanes

The series of papers mostly uses a model structure of a steady hurricane traveling over the open sea. Although the model structure is simplified, it is close to that exposed in many textbooks and papers. The hurricane is viewed as a solitary vertical air vortex, rotating near the bottom in the cyclonic direction. It vertically structured as a central "eye" of radius $r_i \sim 20$km, surrounded by the "eye wall" (EW) with external radius $r_e$ of ~30-40km, and outer region (Fig.1.1). It is structured as consisting of two major layers, the hurricane boundary layer (HBL) at the bottom with the vertical thickness $h_b$, and (almost) adiabatic layer above it where the external radius $r_e$ varies with height $z$. Two relatively thin sub-layers are also important in the HBL. The bottom, turbulent sub-layer located near the air/sea interface, and the upper, condensation sub-layer, restricting the HBL from the top.

The air within the eye and EW performs intense rotation, whose peak is achieved at the outer boundary $r_e$ of EW. In the external region of hurricane, outside the EW, the rotational intensity decreases to zero (relative to the Earth rotation) at the external (ambient) radius $r_a$ of hurricane. Typically, the external radius of hurricane $r_a$ is about 400-600km. According to observations, an upward directed vertical component of airflow of a relatively low intensity is almost exclusively contained within the EW region. Therefore in our simplified modeling it is assumed that the vertical airflow exists only in the hurricane EW. This rotating and ascending airflow in adiabatic layer is called below the *EW jet*. An intense radial airflow happens only at the bottom and upper layers of the hurricane, however, in opposite directions: inwards at the bottom and outwards at the top of hurricane. The entire hurricane height is up to 20-30*km*. One should also mention that two variables typically measured in hurricanes, radial profiles of wind speed (at the upper part of boundary layer) and surface pressure are currently described by empirical relations.

A hurricane could exist only when the temperature in the ascending EW jet is higher than the "ambient" surface temperature $T_a^0$ ( $= T_a|_{z=0}$) far away from the hurricane. To maintain this temperature it is assumed that there exists an environmental, horizontal near-sea warm air band with the temperature $T_+$, which supplies a relatively hot air to the hurricane boundary layer. The geometry of the warm air band is simplistically modeled as a parallelepiped-like path of height $h_b$ and width $H$. It is assumed that

$$H \approx 2r_e. \qquad (1.1)$$

Relation (1.1) means a hypothesis of absorption, delivering fresh air in the traveling hurricane. It also formally means that the hurricane adjusts the bottom EW size to that of the warm air band size to optimize the mass and heat transfer from the environment to the hurricane.

We now briefly discuss the basic physical processes happened in matured hurricanes traveling over open seas. In the most cases this travel as well as all the processes within hurricane can be considered as stationary or quasi-stationary, i.e. slowly changing with time.

Many papers have documented the conservation of angular momentum *M* in the upper layers of hurricane. Nevertheless, *M* seems to be growing in radial direction in the hurricane boundary layer.

Hurricanes travel horizontally over the open sea interacting with environment. Various physical processes occur in the boundary layer due to this interaction. Apart from direct air-sea dynamic interaction at the air/sea interface and evaporation, the heat/mass exchange between the hurricane and ambient warm air band is also important. The active heat, moisture and the latent heat are then transported upwards from the HBL via EW jet. The height of HBL is limited by the intense condensation process, assumed in this modeling being confined in a thin zone, as in the case of slow combustion [24]. The condensation makes the airflow two-phase with occurrence of water droplets spraying upwards and outside the EW jet. It causes the formation of spiral rain bands, layered clouds and rainfall from them. Although the rainfall also causes the downward vertical air motion outside the EW, we neglect the contribution of this motion in the dynamics of hurricane. Above HBL, the air flows will roughly be considered as adiabatic, ignoring additional condensation and its contribution to the spiral rain bands.

Important consequence of the above modeling is the axial symmetry of airflows above the boundary layer. Nevertheless, the airflow in the HBL cannot be considered as axially symmetric because of the horizontal motion of hurricane. So axially symmetric model of HBL developed in paper 3 could be considered only as a rude approximation.

Additionally, hurricane generates oceanic waves and creates some underwater oceanic currents. It also seems reasonable to assume in the modeling HBL that the rainfall balances the evaporation from the oceanic surface underlying the hurricane. This approximate balance will yield a constant salinity level in the oceanic boundary layer.

## 1.3. Preliminary

This Section systematizes well known thermodynamic and aerodynamic equations, used in the following papers of series.

***Thermodynamics of humid air*** [25]

Far away from the tropical cyclone the atmosphere is assumed to be horizontally homogeneous with distributed vertically ambient density $\rho_a$ pressure $p_a$ and temperature $T_a$. These distributions are described in [25] by the equilibrium equations within the thermodynamics of ideal gas with the entropy $S$, internal energy $U$, enthalpy $E$ and free Helmholtz energy $F$ defined as:

$$\Delta S = c_p \ln T - R \ln p + mL_v / T = c_v \ln T - R \ln \rho + mL_v / T$$
$$\Delta U = c_v T, \ E = c_p T + mL_v, \ \Delta F = \Delta U - T\Delta S = RT \ln \rho - mL_v . \tag{1.2}$$

Here $T$ is temperature, $p$ is pressure, $\rho$ is density, $m$ is specific humidity, $L_v$ is specific latent heat, $R = c_p - c_v = R_u / M_a$ is the specific air gas constant, $M_a$ is the molar mass of air, $R_u$ is the universal gas constant, $c_p$ and $c_v$ are the heat capacities of air at constant pressure and constant volume, respectively. Due to (1.2), the state equation for the gas in equilibrium has the familiar form:

$$p = \rho RT . \tag{1.3}$$

We consider the single-phase description for the humid air with the density $\rho$, gas constant $R$ and heat capacity $c_p$ depending on the densities of dry air $\rho_d$, the density of vapor $\rho_v$, and the small enough moisture concentration $m$ as:

$$\rho = m\rho_v + (1-m)\rho_d, \ R = R_d(1-m) + R_v m, \ c_p = c_{pd}(1-m) + c_w m . \tag{1.4}$$

Here $R_d$ and $R_v$ are the gas constants for dry air and vapor, $c_{pd}$ is the heat capacities at constant pressure of dry air and $c_w$ is the heat capacity of water. Usually the dew temperature $T_d$ is increasing function of moisture concentration. Thus for small enough variations of moisture $\Delta T_d \sim \Delta m$. In the following rough enough modeling we will also use the virtual temperature $T_v = T_d + \Delta T_d$ as a thermodynamic field variable, which is no more than $(2-3)^0 C$ higher than the real temperature $T$.

Additionally, there is the equilibrium equation describing the vertical pressure distribution in the horizontally homogeneous atmosphere:

$$dp_a / dz = -\rho_a g . \tag{1.5}$$

We consider in this paper the adiabatic description of air when the relation (1.3) is represented as $p_a = p_a^0 (\rho_a / \rho_a^0)^\gamma$, with the adiabatic index $\gamma = c_p / c_v \approx 1.4$. Then equations (1.3) and (1.5) yield:

$$\rho_a(z)/\rho_a^0 = \psi^{\frac{1}{\gamma-1}}, \ p_a(z)/p_a^0 = \psi^{\frac{\gamma}{\gamma-1}}, \ T_a(z)/T_a^0 = \psi, \ \psi(z) \equiv 1 - \frac{(\gamma-1)gz}{\gamma RT_a^0} . \tag{1.6}$$

Here $p_a^0 = \rho_a^0 R T_a^0$ and $p_a^0, \rho_a^0$ and $T_a^0$ are ambient thermodynamic parameters at the surface. These values of parameters will be furthermore used in (1.2) as the standard ones. Hereafter the ambient temperature $T_a^0$ is considered as real, i.e. not virtual.

*Aerodynamic equations for air flows* [26]

The following modeling uses simplified aerodynamic equations for hurricane. To clearly expose below the model assumptions we remind here the formulation of well known equations for ideal compressible gas. We consider axially symmetric, rotating air flows around a vertical $z$ - axis (directed upward) as a finite disturbance of static distributions (1.6 a, b) for density, pressure and temperature. A frame of reference below consists of a cylindrical coordinate system slowly moving in a horizontal direction. In this coordinate system, the motions of air relative to the local Earth rotation can be described by the aerodynamic equations written in the divergent form as:

$$\partial_t(\rho r) + \partial_r(\rho r u_r) + \partial_z(\rho r u_z) = 0 \tag{1.7}$$

$$\partial_t(\rho r M) + \partial_r(\rho r u_r M) + \partial_z(\rho r u_z M) = 0, \quad M = r(u_\varphi + \Omega r) = r^2(\omega + \Omega) \tag{1.8}$$

$$\partial_t(\rho r u_z) + \partial_r(\rho r u_r u_z) + \partial_z(\rho r u_z^2) = r[\partial_z(p_a - p) + g(\rho_a - \rho)] \tag{1.9}$$

$$\partial_t(\rho r u_r) + \partial_r(\rho r u_r^2) + \partial_z(\rho r u_r u_z) + r \partial_r p = \rho \chi r^2, \quad \chi = (\omega - \Omega)^2 \tag{1.10}$$

Equation (1.7) describes the continuity (mass conservation), equations (1.8)-(1-10) present the time evolution of momentum components in radial $r$, tangential $\varphi$, and vertical $z$ directions. There $u_r, u_\varphi$ and $u_z$ and are the respective velocity components, $M$ is the angular momentum (per mass unit) of absolute rotation, $\omega$ is the angular velocity of relative rotation, and $\Omega (= \Omega_e \sin \beta)$ is the local Earth rotation on $\beta$ - plane where $\Omega_e$ is the absolute Earth rotation. The additional centrifugal term $\Omega^2$ of the local Earth rotation [26], is included in (1.10) along with common Coriolis term. In equations (1.6) and (1.7)-(1.10), $p_a, \rho_a$ and $p, \rho$ are the ambient and actual pressures and densities, which should be related by specific constitutive rules, as e.g. in the adiabatic case $p_a = \rho R T_a$, $p = \rho R T$.

Equations (1.8-1.10) can also be rewritten with the aid of (1.7) in equivalent non-divergent form. E.g. equations (1.8) for conservation of angular moment and (1.9) for balance of radial momentum can be rewritten in the familiar forms as:

$$\partial_t M + u_r \partial_r M + u_z \partial_z M = 0 \qquad (1.8a)$$

$$\partial_t u_r + \partial_r u_r^2 / 2 + u_z \partial_z u_r + \rho^{-1} \partial_r p = \chi r . \qquad (1.9a)$$

In the above axisymmetric case it is sometimes convenient to introduce, instead of vertical and radial velocity components, the "mass" stream function $\Psi(r, z)$ by the familiar relations:

$$\rho r u_r = -\partial_z \Psi, \qquad \rho r u_z = \partial_r \Psi. \qquad (1.11)$$

Then in steady axisymmetric case equations (1.8) or (1.8a) yield the first integral:

$$M = \Phi(\Psi). \qquad (1.12)$$

Here $\Phi$ is an arbitrary function of $\Psi$. The proof immediately follows upon substitution (1.11) into the stationary version of (1.8a) and reducing this equation to

$$\mathbf{J}_{z,r}(M, \Psi) \equiv \partial_z M \cdot \partial_r \Psi - \partial_r M \cdot \partial_z \Psi = 0.$$

Here $\mathbf{J}_{z,r}$ is the Jacobean of two functions with respect of $z, r$ variables.

It should be noted that parameters which may characterize the function $\Phi$ in (1.12) must be space independent. E.g. in case of linear dependence (1.12) $M = \alpha_1 \Psi + \alpha_2$, parameters $\alpha_1$ and $\alpha_2$ are constants (or constant functionals of $\Psi$). Existence of such a first integral is helpful in aerodynamic modeling of stationary hurricane dynamics.

The same approach is applicable to the large scale heat and mass transport processes in hurricanes with dominant advection mechanism. E.g. for advective heat transport in stationary axisymmetric case described by equation,

$$\partial_r (\rho c_p r u_r T) + \partial_z (\rho c_p r u_z T) = 0,$$

where $T$ is the temperature and $c_p$ is the heat capacity at constant pressure, there is another first integral,

$$T = \Phi_1(\Psi).  \qquad (1.13)$$

Here $\Phi_1$ is another arbitrary function of $\Psi$. Formulas (1.12) and (1.13) also show that in this type of air flow, all three variables $\Psi, M$ and $T$ are functionally related.

## 1.4. Principles of horizontal travel of hurricanes

Three factors affect the translational horizontal travel of hurricane relative its center of mass: (i) stirring or "sailing" wind, (ii) a tendency for accepting more hot and humid air from environment ("affinity" factor), and (iii) the Coriolis force. The sailing wind $\underline{w}_s$ with common values $w_s \approx 2$m/s is a primary factor considered in many books and papers as synoptically related large scale air current in which "the hurricanes are imbedded" [6]. Another, affinity (or "fueling") factor is related to the necessity to maintain considerably higher temperature level in the hurricane EW jet as compared with the respective ambient level. This factor is similar to tendency of hurricane to accept the maximal potential vorticity postulated in many papers (e.g. see Ref.[21]). Thus when the sailing wind and Coriolis forces are negligible, the hurricane horizontal motion can be described by the following "affinity" principle: Without the sailing wind and neglecting the Coriolis force the hurricane travels in the direction of maximal surface air temperature $T_+$, which may exist at the external radius $r_{e0}$ of the hurricane EW jet bottom.

The affinity principle of hurricane motion is in accord with the well known observations that tropical cyclones are moving along the warm currents, being completely absent in South Atlantic and Eastern South Pacific (e.g. see Ref.[2], p.48). This is because no warm ocean currents exist in these areas that cause the warm near-water air bands to exist. Note that in the Western North Atlantic where hurricanes tend to propagate in Northern direction (in general close to the Gulf Stream), the Coriolis force directed to the West, reinforces the hurricanes to move in NNW direction.

Consider now the hurricane EW jet approaching the maximal temperature warm air band, when stirring wind and Coriolis force are negligible. This steady or quasi-steady temperature field outside the hurricane caused by long-term synoptic variations is assumed to be known. Let $\Gamma$ be a curve of max $T_+$ with $x$ being a distance counted off along the curve $\Gamma$ in the direction of temperature increase, and $\underline{y} = \underline{y}(x)$ be a tangent unit vector along the curve $\Gamma$, directed toward the side of $x$ growing. Then the affinity force is $\sim \underline{\nabla}_\Gamma T$. According to the affinity principle the hurricane EW jet preferably moves along the curve $\Gamma$ with a speed $w_a$, where the temperature field $T(x)$ is considered as known. Then the total velocity $\underline{u}_h$ presenting the sailing and affinity components of horizontal motions of hurricane, along with the Coriolis force, is described as

$$d/dt(\underline{u}_h - \underline{w}_s - w_a \underline{y}) = 2\underline{\Omega} \times \underline{u}_h. \qquad (1.14)$$

Here $\underline{\Omega}$ is the local angular velocity of Earth on $\beta$ plane. Equation (1.14) has kinematic form because each term there has approximately equal mass multiplier of the hurricane column. Using common thermodynamic relations in the adiabatic layer of hurricane (see the Section 1.3) one could show that a partial motion of tropical cyclones in affinity direction causes the increase in the EW jet entropy or decrease in its free energy. It means that this type of hurricane travel makes the tropical cyclone system more thermodynamically stable.

## References

1. Gordon E. Dunn, Tropical Cyclones, in: *Compendium of Meteorology*, Am. Meteorol. Soc., pp. 887-901 (1951).
2. R. A. Anthes, *Tropical Cyclones, Their Evolution, Structure and Effects*, Am. Met. Soc., Science Press, Ephrata, PA (1982).
3. S.A. Hsu, *Coastal Meteorology*, Acad. Press, New York (1988).
4. W.R. Cotton and R.A. Anthes, *Storms and Cloud Dynamics*, Acad. Press, San Diego (1989)
5. A. Ogawa, *Vortex Flow*, CRC Press. INC., p. 311 (1993)
6. K. A. Emanuel, *Divine Wind*, Oxford Univ. Press (2005)
7. K. A. Emanuel, Thermodynamic control of hurricane intensity, *Nature*, **401**, 665-669 (1999).
8. K. A. Emanuel, The behavior of simple hurricane model using a convective scheme based on subcloud –layer entropy equilibrium, *J. Atmos. Sci.*, **52**, 3959-3968 (1995).
9. K. A. Emanuel, An air-sea interaction theory for tropical cyclones. Part I. *J. Atmos. Sci.*, **42**, 1062-1071 (1983).
10. G.J. Holland, The maximal potential intensity of tropical cyclones, J. Atmos. Sci., 54, 2519-2541 (1997).

11. G.J. Holland and R.T. Merrill, On the dynamics of tropical cyclone structural changes, *Quart. J. Roy. Meteorol. Sci*. **110**, 723-745 (1984).
12. M. Bister and K.A. Emanuel, Dissipative heating and hurricane intensity, *Meteorol. Atmos. Phys.*, **65**, 233-240 (1998).
13. M.T. Montgomery, V.A. Vladimirov and P.V. Denissenko, An experimental study on hurricane mesovortices, *J. Fluid Mech*., **471,** 1-32 (2002).
14. J. Lighthill, Ocean spray and the thermodynamics of tropical cyclones, *J. Engn. Math,* **35**, 11- 42 (1999).
15. G. I. Barenblatt, A.J. Chorin and V.M. Prostokishin, A note concerning the Lighthill Sandwich Model of tropical cyclones, *Proc. Nat. Acad. Sci. (PNAS),* **102**, 11148-11150 (2005).
16. B.A. Kagan, *Ocean-Atmospheric Interaction and Climate Modeling*, Cambridge University Press, New York (1995)
17. K.V. Ooyama, Numerical simulations of life cycle of tropical cyclones, *J. Atmos. Sci*., **26**, 3-40 (1983).
18. K.V. Ooyama, Conceptual evolution of the theory and modeling of the tropical cyclone. *J. Met. Soc. Japan*, **60**, 369-379 (1982).
19. K. A. Emanuel, The finite amplitude nature of tropical cyclogenesis, *J. Atmos. Sci*., **46**, 3431-3456 (1989).
20. K. A. Emanuel, Some aspects of hurricane inner-core dynmics and energetics, *J. Atmos. Sci*., **54**, 1014-1026 (1997).
21. Y. Liu, D.-L. Zhang and M.K. Yao, A multi-scale numerical study of hurricane Andrew (1992). Part I: Explicit simulation and verification, *Month. Weather Rep*., **125,** 3073-3093 (1997).
**22.** Y. Wang and C.-C. Wu, Current understanding of tropical cyclone structure and intensity changes – a review, *Meteorol. Atmos. Phys.,* **87,** 257-278 (2004).
23. J. C. Chang, The physics of tropical cyclone motion, *Ann. Rev. of Fluid Mech.* **0115**, 99-128 (2005).
24. L.D. Landau and E.M. Lifshitz, "*Theoretical Physics.* Vol.IV*: Fluid Mechanics",* Sect.132 (in Russian). Nauka, Moscow (1986).
25. J.A. Curry and P.J. Webster, *Thermodynamics of Atmospheres and Oceans*, Acad. Press, London ((1999).
26. G.K. Batchelor, "*An Introduction to Fluid Dynamics*", Cambridge Univ. Press, Cambridge (1970).

## Acknowledgment

The author is thankful to Dr. A. Benilov (Stevens Institute of Technology, NJ) for extensive and productive discussions of all the papers of the series. A lot of thanks are also given to my former PhD Student A. Gagov for valuable help in calculations and graphics.

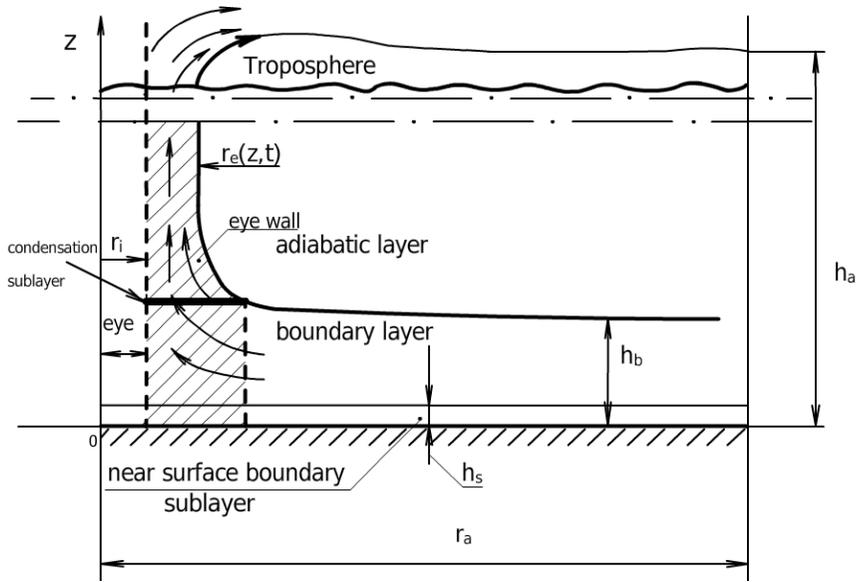

Fig.1.1: Schematic structure of hurricane